# Magnetic order induced chiral phonons in a ferromagnetic Weyl semimetal

Mengqian Che[1], Jinxuan Liang[1], Yunpeng Cui[1], Hao Li[2], Bingru Lu[2], Wenbo Sang[2], Xiang Li[2], Xuebin Dong[3], Le Zhao[1], Shuai Zhang[4], Tao Sun[3], Wanjun Jiang[1,5], Enke Liu[3], Feng Jin[3], Tiantian Zhang[4,*], Luyi Yang[1,5,*]

[1]State Key Laboratory of Low Dimensional Quantum Physics and Department of Physics, Tsinghua University, Beijing, 100084, China

[2]School of Physical Science and Technology, Lanzhou University, Lanzhou 730000, China

[3]Beijing National Laboratory for Condensed Matter Physics and Institute of Physics, Chinese Academy of Sciences, Beijing 100190, China

[4]Institute of Theoretical Physics, Chinese Academy of Sciences, Beijing 100190, China

[5]Frontier Science Center for Quantum Information, Beijing 100084, China

*Emails: ttzhang@itp.ac.cn; luyi-yang@mail.tsinghua.edu.cn

## Abstract

**Chiral phonons are vibrational modes in a crystal that possess a well-defined handedness or chirality, typically found in materials that lack inversion symmetry. Here we report the discovery of chiral phonon modes in the kagome ferromagnetic Weyl semimetal $Co_3Sn_2S_2$, a material that preserves inversion symmetry but breaks time-reversal symmetry. Using helicity-resolved magneto-Raman spectroscopy, we observe the spontaneous splitting of the doubly degenerate in-plane $E_g$ modes into two distinct chiral phonon modes of opposite helicity when the sample is zero-field cooled below the Curie temperature, in the absence of an external magnetic field. As we sweep the out-of-plane magnetic field, this $E_g$ phonon splitting exhibits a well-defined hysteresis loop directly correlated with the material's magnetization. The observed spontaneous splitting reaches up to 1.27 cm$^{-1}$ at low temperatures, progressively diminishes with increasing temperature, and completely vanishes near the Curie temperature. Our findings highlight the role of the magnetic order in inducing chiral phonons, paving the way for novel methods to manipulate chiral phonons through magnetization and vice versa. Additionally, our work introduces new possibilities for controlling chiral Weyl fermions using chiral phonons.**

The interplay between magnetism and chiral phonons is a captivating frontier in condensed matter physics, materials science, and spintronics. Chiral phonons are quantized modes of lattice vibrations characterized by rotational motion [1]. These intriguing phonons naturally arise in non-centrosymmetric materials, such as monolayer transition-metal dichalcogenides like $WSe_2$ [2], and chiral crystals including quartz, α-HgS and Te [3-5]. In crystals that break time-reversal symmetry, magnons can selectively couple with phonons when they are in resonance, giving rise to the formation of chiral magnon polarons [6, 7]. Remarkably, in a diverse range of materials – including semiconductors such as PbTe, Dirac semimetals like $Cd_3As_2$, 4$f$ rare-earth halide paramagnets such as $CeF_3$ and $CeCl_3$, as well as 3$d$ transition-metal oxide magnets like $Fe_2Mo_3O_8$ and $CoTiO_3$ – the application of a magnetic field can induce substantial chiral phonon Zeeman splitting [8-14]. This phenomenon uncovers a surprisingly large phonon magnetic moment – orders of magnitude greater than what would be expected from ionic cyclotron motions alone. This giant effective chiral phonon magnetic moment offers a novel mechanism for controlling the electronic and magnetic properties of materials [15-19]. Beyond the influence of external magnetic fields, chiral phonons can also be effectively manipulated through resonant pumping techniques using mid-infrared or terahertz pulses [10, 20-24], enabling ultrafast control of magnetic order.

However, while progress has been made in the field of chiral phonons in magnetic materials [6, 11-13, 23] the control of chiral phonons has predominantly been demonstrated through the application of external magnetic fields. Research exploring the intricate correlations between magnetic order and chiral phonons is still at an early stage. Understanding these correlations could lead to deep insights into the fundamental physics of spin-chiral phonon coupling, as well as paving the way for innovative applications in spintronics and quantum technology.

In this work, we observe magnetic order-induced chiral phonons and their spontaneous splitting in the ferromagnetic Weyl semimetal $Co_3Sn_2S_2$ using helicity-resolved magneto-Raman scattering spectroscopy. Below the ferromagnetic transition temperature, a notable splitting occurs between the two enantiomeric $E_g$ phonon modes, driven by spontaneous magnetization even in the absence of an externally applied magnetic field. The chiral splitting exhibits magnitudes of up to 1.27 cm$^{-1}$ at low temperatures, gradually diminishes with increasing temperature, and ceases entirely near the Curie temperature. The splitting of chiral phonons of opposite helicity correlates with magnetization, but exhibits minimal dependence on the magnetic field, indicating that it arises from strong spin-phonon interactions in $Co_3Sn_2S_2$. To the best of our knowledge, this is the first demonstration of chiral phonons and their spontaneous splitting in a centrosymmetric ferromagnet. It is the magnetic order that breaks time-reversal symmetry and gives rise to these chiral phonon modes. Our findings open the avenues for controlling chiral phonons with magnetization and vice versa.

$Co_3Sn_2S_2$ is a topological ferromagnet exhibiting a unique interplay between its electronic wavefunction topology and magnetic spin configuration. This interaction results in several intriguing properties, including

Weyl points [25-27], Fermi arcs [28, 29], a colossal anomalous Hall effect [26, 30, 31], a pronounced magneto-optical response [32, 33], and most recently, the discovery of terahertz-frequency magnons [34]. These properties collectively make $Co_3Sn_2S_2$ a rich platform for exploring novel quantum phenomena and potential applications in spintronics and novel electronic devices.

Figure 1a illustrates the crystal structure of $Co_3Sn_2S_2$, with the primitive unit cell highlighted. In this structure, Co atoms form the kagome lattice with a Sn atom located at each hexagon's center. The $Co_3Sn$ layer is intercalated between two hexagonal S layers, which are further sandwiched between two hexagonal Sn layers. $Co_3Sn_2S_2$ crystallizes in the centrosymmetric space group $R\bar{3}m$ (No. 166) with a point group of $\bar{3}m$. Below the Curie temperature of ~175 K, the ground state is ferromagnetic, with the spins (represented by red arrows in Fig. 1a) aligned along the easy axis of magnetization, which corresponds to the $c$ direction of the crystal. Notably, the space group symmetry of the ferromagnetic phase remains unchanged from that of the paramagnetic phase, while the magnetic group of the electronic system changes to $\bar{3}m'$ [35]. However, when considering the symmetry of phonons, only the lattice symmetry is typically considered. It is the spin-lattice interaction in the ferromagnetic state that breaks time-reversal symmetry in the phonon system.

The primitive unit cell of $Co_3Sn_2S_2$ contains three Co atoms, two Sn atoms, and two S atoms, totaling seven atoms. As a result, the material has 21 phonon modes: 3 acoustic and 18 optical. At the Γ point, the phonon modes can be categorized into irreducible representations as follows: $\Gamma = A_{1g} + E_g + A_{1u} + 5A_{2u} + 6E_u$. Among these 21 modes, 3 optical modes are Raman active: one with nondegenerate $A_{1g}$ symmetry and two with doubly degenerate $E_g$ symmetry. The remaining 18 modes, including the acoustic ones, are infrared active. The $A$ and $E$ symmetry modes represent out-of-plane and in-plane vibrations, respectively. Figures 1b and 1c display the first Brillouin zone and the phonon dispersion along the Γ-T ($k_z$) direction as calculated by density functional theory (DFT), with the Raman active modes highlighted in red. Detailed dispersion relations are provided in Fig. S2. Figures 1d and 1e illustrate atomic motions of the Raman active modes. The $A_{1g}$ mode corresponds to the movement of S atoms along the $c$ axis, while the $E_g$ modes involve S atoms moving within the $ab$ plane. Notably, these $E_g$ modes can be decomposed into a pair of phonons exhibiting opposite circular motions, as shown in Fig. 1e.

We performed helicity-resolved magneto-Raman scattering measurements in the backscattering geometry with the incident and scattered light normal to the $ab$ plane (along the $c$ axis of the crystal). Therefore, the phonon momentum involved in the Raman process is along the $c$ axis of the crystal, which is perpendicular to the motion of the $E_g$ mode. The sample was mounted in a superconducting magnet system with the magnetic field applied in the out-of-plane direction (along the $c$ axis of the crystal). Further details of the setup can be found in the Methods section in the Supplementary Material. Unless otherwise stated, most data were taken with a He-Ne laser at 632.8 nm. The corresponding scattered phonon wavevector is given by $k_{\rm ph} = \frac{4\pi}{\lambda}n = 5.7 \times 10^5$ cm$^{-1}$, which is about 1/120 of the first Brillouin zone size along the $k_z$ direction, where $\lambda$ is the wavelength of the incident light in vacuum and $n = 2.87$ is the corresponding real part of the complex index of refraction [36]. We conducted systematic measurements on two $Co_3Sn_2S_2$ samples as a function of light polarization, field and temperature, with both samples yielding consistent results.

We begin by presenting the results at room temperature. As shown in Fig. 2a, the Stokes Raman spectra obtained using co-circularly (*RR* or *LL*) and cross-circularly (*RL* or *LR*) polarized incident and scattered photons were used to identify the symmetries of the Raman active modes, where *R/L* stands for right/left circularly polarized light. The $A_{1g}$ mode appears only in the co-circular polarization. In contrast, the $E_g$ modes are detectable only in the cross-circularly polarized configuration. These results agree with the Raman selection rules (Supplementary Note 3). The peak frequencies (385.07 and 288.52 cm$^{-1}$) match well with our DFT calculations (Fig. 1c) and previous studies [37, 38].

In the helicity-resolved Raman scattering process, in addition to the conservation of energy and momentum, the pseudo-angular momentum is also conserved [39-41]: $l_{\rm i} = l_{\rm s} + l_{\rm ph}$, where $l_{\rm i}$, $l_{\rm s}$, and $l_{\rm ph}$ denote the (pseudo-)angular momenta of the incident photon, scattered photon, and emitted phonon, respectively, in the Stokes Raman scattering process. Since $l_{\rm ph} = 0$ for the $A_{1g}$ mode, it appears only in the co-circular polarization, where $l_{\rm i} = l_{\rm s}$. In contrast, for the two $E_g$ modes, $l_{\rm ph} = \pm 1$, allowing for the following the conservation rules: $l_{\rm i}(-1) = l_{\rm s}(+1) + l_{\rm ph}(+1)$ *mod* 3 or $l_{\rm i}(+1) = l_{\rm s}(-1) + l_{\rm ph}(-1)$ *mod* 3. The *mod* 3 is due to the three-fold rotational symmetry of the crystal along the *c* axis and the Umklapp process in Raman scattering. Therefore, these two $E_g$ modes are chiral phonon modes with $l_{\rm ph} = \pm 1$, and they are detectable only in the cross-circularly polarized configuration. At room temperature, these two $E_g$ modes are degenerate due to the presence of time-reversal symmetry, as expected.

Surprisingly, as we zero-field cooled the sample to 2 K, the doubly-degenerate $E_g$ modes spontaneously split into two modes centered around 293.80 and 295.07 cm$^{-1}$, each with distinct peak intensities and linewidths (Fig. 2b). Each of these $E_g$ modes is selectively detectable in either the *RL* or *LR* configuration, respectively. To the best of our knowledge, this is the first observation of energy splitting between chiral phonons triggered by spontaneous magnetization in a ferromagnet. Although chiral phonons have been observed in other magnetic materials [12, 13, 42], like $Fe_2Mo_3O_8$, $CoTiO_3$ and $CrBr_3$, external magnetic fields are required. In contrast, the energy splitting observed in $Co_3Sn_2S_2$ is unexpectedly large even in the absence of external magnetic fields, reaching 1.27 cm$^{-1}$ = 38.1 GHz, surpassing that in chiral crystals of α-HgS and Te. Notably, the chiral phonons identified here possess nonzero helicity, as their wavevector is along the *c* axis while vibrating within the *ab* plane. This differs from those observed in 2D materials, where chirality is not well-defined.

We find that the energy splitting, defined as $f_{LR} - f_{RL}$ with $f_{LR}/f_{RL}$ being the center frequencies for the *LR/RL* configurations, correlates with the amplitude and direction of magnetization. The splitting changes sign when the direction of magnetization reverses in response to an out-of-plane magnetic field (Fig. 2c). Sweeping the applied field reveals that both center frequencies exhibit a well-defined hysteresis loop (Fig. 2d). Figure 2e illustrates the splitting as a function of the applied field, which exhibits a square-like hysteresis loop. In contrast, for a fixed magnetization direction, the splitting shows only a slight variation with the applied field, remaining unchanged up to 6 T (Fig. 2d and additional data in Figs. S5 and S6).

This observation indicates that the splitting of chiral phonons is directly linked to magnetization rather than the applied magnetic field. In $Co_3Sn_2S_2$, a hard ferromagnet with strong easy-axis magnetic anisotropy [26, 34, 43], the magnetic moments spontaneously align within domains (typically larger than 20 μm, Fig. S7, [33]) that are much larger than the probe laser's focal spot (~2 μm) at zero field. As a result, the chiral phonon splitting remains unchanged with the external field, provided the magnetization is not flipped by the field. This behavior contrasts sharply with previous findings [11-14] in the paramagnet $CeCl_3$ and the antiferromagnets $Fe_2Mo_3O_8$ and $CoTiO_3$, where the magnetic moments are gradually aligned by the external field, displaying a linear relationship with the magnetic field at low fields.

The Raman shifts for $E_g$ obtained above and below the Curie temperature reveal the degeneracy and splitting of chiral phonon modes influenced by the magnetic order. Additionally, reversing the direction of the ferromagnetic order also reverses the splitting direction of the chiral phonon modes. These results strongly suggest that the splitting of the two chiral phonon modes in $Co_3Sn_2S_2$ is influenced by spin-phonon coupling, which can be explained through two approaches: (1) by considering how the phonon force constant is modified by magnetization [44], and (2) by examining the effect of the nuclear/molecular Berry curvature (MBC) on the lattice in the presence of ferromagnetic order [45-50].

The first case is based on symmetry analysis (see Supplementary Note 6). In a uniaxial crystal, to the linear order, magnetization (or an applied magnetic field) along the optical axis (e.g., *c* axis in $Co_3Sn_2S_2$) alters the off-diagonal elements of the $E_g$ force constant: $K_{xy}(M_z, H_z) = K_{xy}^{(0)} + i(K_M M_z + K_H H_z)$ and $K_{yx}(M_z, H_z) = K_{yx}^{(0)} - i(K_M M_z + K_H H_z)$, where $K_{xy(yx)}^{(0)}$ is the unperturbed force constant, and $K_{M(H)}$ is the coupling constant to the magnetization $M_z$ (external field $H_z$). This modification leads to the splitting of the doubly degenerate $E_g$ modes and the chiral phonons become the eigenmodes. The splitting should be linear in both $M_z$ and $H_z$. This phenomenon is also called "phonon Zeeman effect" [51]. The absence of field dependence on phonon splitting (from the coercive field up to 6 T) suggests that the coupling of the chiral phonons with magnetization is significantly stronger than their coupling with the applied field.

The second case involves the MBC. Under the Born-Oppenheimer approximation, the electronic ground state evolves adiabatically with the lattice and accumulates a geometrical Berry phase, which in turn influences the lattice dynamics as a gauge field in the lattice Hamiltonian [45-50, 52]. The kinetic energy changes to $\sum_\alpha \frac{(p_\alpha - A_\alpha)^2}{2m_\alpha}$, where $m_\alpha$ is the mass, $p_\alpha$ is the canonical momentum, and $A_\alpha = i\langle \Psi_e | \nabla_\alpha | \Psi_e \rangle$ is the nuclear/molecular Berry connection with $\Psi_e$ being the electron ground state wavefunction and $i$ being the imaginary unit. The equation of motion for the lattice is given by $m_\alpha \ddot{q}_\alpha = -\sum_\beta K_{\alpha\beta} q_\beta + \sum_\beta G_{\alpha\beta} \dot{q}_\beta$, where $q_\beta$ is the lattice displacement, $K_{\alpha\beta}$ is the force constant, and $G_{\alpha\beta}$ is the MBC in the parameter space of nuclear coordinates, which is related to the Berry connection by $A_\alpha = \sum_\beta G_{\alpha\beta} q_\beta$. The modified equation of motion shows that the MBC acts as a nonlocal effective magnetic field on the lattice, and it becomes nontrivial when time-reversal symmetry is broken. Since $Co_3Sn_2S_2$ is a Weyl semimetal without time-reversal symmetry [26, 31, 53], we expect the MBC to have a significant effect on the chiral phonon splitting in the vicinity of the Brillouin zone center. Reversing the magnetization direction would also flip the sign of the MBC, resulting in opposite energy differences between *LR* and *RL* measurements. A calculation of the spin-phonon coupling contributed by the MBC by DFT could provide insight into the origin of this large splitting, though such an analysis is beyond the scope of this study.

Previous experimental and theoretical studies [6, 48] have demonstrated that when chiral phonons are resonant with magnons, their interaction can lead to hybridized magnon polarons. In $Co_3Sn_2S_2$, the acoustic magnon frequency at the Γ point is measured to be ~0.6 THz at low temperatures [34], corresponding to a spin wave gap of 2.3 meV [35, 54], which is an order of magnitude smaller than the chiral phonon frequencies. In addition, Raman-active phonon modes couple with magnons of the same symmetry [48], whereas in $Co_3Sn_2S_2$ the magnons modes are infrared-active. Therefore, the coupling between phonons and magnons is not expected to play a significant role here. Instead, it is the spin-phonon coupling that generates the chiral phonons.

We next investigate the temperature dependence of the chiral phonon modes. As shown in Figs. 3a, the $E_g$ modes soften, and the splitting between them diminishes with increasing temperature. We plot the results of the $A_{1g}$ mode in Fig. S8. We fit the Raman shift with a Lorentzian function: $\frac{A_{\alpha\beta}}{\pi} \frac{\gamma_{\alpha\beta}}{(f - f_{\alpha\beta})^2 + \gamma_{\alpha\beta}^2} + C$, where $\alpha\beta$ denotes either *RL* or *LR*, and $f_{\alpha\beta}$, $\gamma_{\alpha\beta}$, $A_{\alpha\beta}$ and $C$ represent the center frequency, linewidth, amplitude and a constant, respectively. The extracted center frequencies and linewidths are plotted as a function of temperature in Figs. 3b, c. At the base temperature of 2 K, these parameters are distinct for opposite chiral phonon modes, but they converge as the temperature approaches the Curie temperature. As the temperature increases, the shift of the phonon modes to lower frequencies is attributed to anharmonic effects that cause lattice expansion [55]. The amplitudes from the fits are almost the same for both modes (not shown).

Figure 3d illustrates the temperature dependence of the energy splitting ($f_{LR} - f_{RL}$) between the two chiral phonon modes for both samples. With increasing temperature, the splitting diminishes, similar to the trend observed in magnetization (Fig. S4a). However, the splitting does not directly scale with magnetization and vanishes at ~150 K, which is below the Curie temperature of ~175 K. This behavior is partially attributed to the laser heating effect, estimated through modeling [56] to raise the temperature by ~10 K at 150 K. In addition, as temperature goes down, $Co_3Sn_2S_2$ undergoes a phase transition from a nodal ring semimetal to a Weyl semimetal around the Curie temperature [53]. This transition alters the Fermi surface due to changes in both the magnetic moment and lattice constants. Therefore, at lower temperatures, the MBC can be enhanced, leading to a more pronounced chiral phonon splitting. This helps explain the temperature-dependent difference between the chiral phonon splitting and magnetization, highlighting the unique role of the MBC in governing chiral phonon dynamics [45, 49]. Moreover, the chiral phonon splitting is insensitive to the applied field, which would align all the domains in a single out-of-plane direction at a relatively small field of tens to hundreds of mT. Therefore, it is unlikely that the complex spin configurations indicated in Ref. [57] have a significant influence on our results.

Figure 3e plots the lifetimes of the two chiral modes converted from the linewidths in 3c for both samples. The lower-frequency phonon mode has a longer lifetime than that of the higher-frequency one. Phonon decay rates are governed by anharmonic processes, in which a phonon decays into two or more phonons. The simplest case involved here is a near zone-center $E_g$ phonon decaying into two optical phonons, each with half the frequency and opposite momentum. The difference in the lifetimes of the two chiral phonon modes may be due to variations in the available decay channels.

In addition, we performed measurements at three different probe wavelengths (632.8, 532.1 and 473.1 nm) and found that the splitting exhibits minimal wavelength dependence (Fig. 4). This behavior contrasts sharply with that of inversion-breaking chiral crystals like α-HgS and Te [4, 5], where phonon modes with distinct chirality are degenerate at the Γ point and split when the momentum deviates from Γ. Our findings indicate that in $Co_3Sn_2S_2$, the splitting occurs at the Γ point, as a result of the breaking of both the time-reversal symmetry and vertical mirror symmetry. This is also different from materials with inversion symmetry breaking, such as monolayer $WSe_2$, where chiral phonons are found at K and K', which are the high symmetry points of the Brillouin zone edge [2].

In summary, by using helicity-resolved magneto-Raman spectroscopy, we have directly observed magnetic order induced chiral phonons of the $E_g$ modes, along with their spontaneous energy splitting in the ferromagnetic Weyl semimetal $Co_3Sn_2S_2$. Our findings suggest that these chiral phonons can be effectively controlled by the magnetization of the material. In addition to the Raman-active modes observed in this study, infrared-active $E_u$ modes, though not detectable by Raman spectroscopy, may also exhibit chiral phonon modes associated with magnetic order, which can be measured using infrared and terahertz spectroscopy. Conversely, we anticipate that resonantly driving chiral phonons in $Co_3Sn_2S_2$ (on the order of tens of meV) may facilitate the manipulation of the magnetic order itself [10, 20].

Moreover, the energy of chiral phonons is comparable to low-energy quasiparticles in the material, chiral Weyl fermions (~50-60 meV above the Fermi level [26, 29]), raising intriguing possibilities for their interaction. It will be fascinating to explore how these chiral bosons couple directly with chiral fermions, as this coupling could open new avenues for controlling topological states in the material [58]. In addition, strain-induced symmetry breaking in $Co_3Sn_2S_2$ can transform the $A_{1g}$ phonon mode to an $A_1$ mode, inducing a resonance in Raman scattering through the interaction between Weyl fermions and phonons [59]. The $A_{1u}$ phonon mode in $Co_3Sn_2S_2$, identified as a pseudoscalar and unpolarized, can exhibit the chiral magnetic effect under a magnetic field [59, 60]. Furthermore, the mutual control of chiral phonons, Weyl fermions, and magnetic order can be achieved on ultrafast timescales using advanced ultrafast techniques. Such capabilities open up exciting possibilities for the development of next-generation devices where topological, phononic and magnetic properties can be dynamically adjusted, paving the way for advances in fields like topological quantum materials, spintronics and quantum computing.

Note added: We note that recently circular dichroism of infrared-active $E_u$ modes in $Co_3Sn_2S_2$ has also been observed by optical spectroscopy, indicating the possible chirality of these modes [61].

## Acknowledgements

The work was supported by the National Key R&D Program of China (Grants No. 2021YFA1400100 and No. 2020YFA0308800), the National Natural Science Foundation of China (Grants No. 12361141826, No. 12421004 and No. 12074212), and the Beijing Natural Science Foundation (Grants No. Z240006). T. Zhang acknowledges the support from the National Natural Science Foundation of China (Grant Nos. 12374165 and 12047503), and National Key R&D Program of China (Grants No. 2023YFA1407400 and No. 2024YFA1400036). This work was supported by the Synergetic Extreme Condition User Facility (SECUF, https://cstr.cn/31123.02.SECUF).

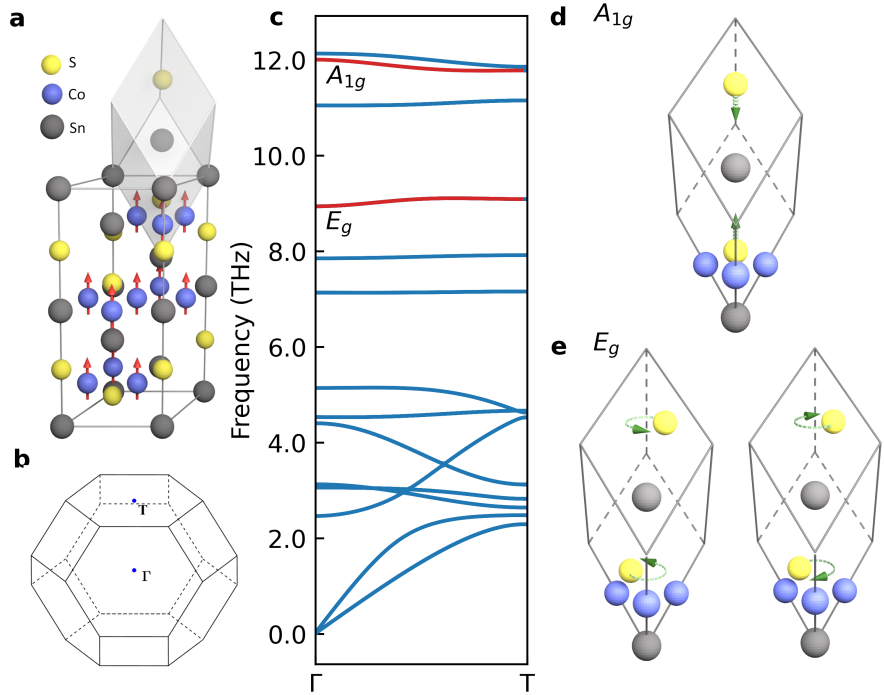

**Fig. 1 Crystal structure and phonon dispersion, and Raman-active phonon modes.** (a) Crystal of structure of $Co_3Sn_2S_2$ with the primitive unit cell outlined. The magnetization is along the *c* axis direction (out of the sample plane) indicated by the red arrows. (b) The first Brillouin zone of the crystal. (c) Phonon dispersion along the Γ-T ($k_z$) direction with the Raman active modes highlighted in red. (d,e) The corresponding atomic motions of the Raman-active modes: in the $A_{1g}$ mode (d), S atoms move along the *c* axis, while in the two enantiomeric $E_g$ modes (e), S atoms move in circular motions within the *ab* plane.

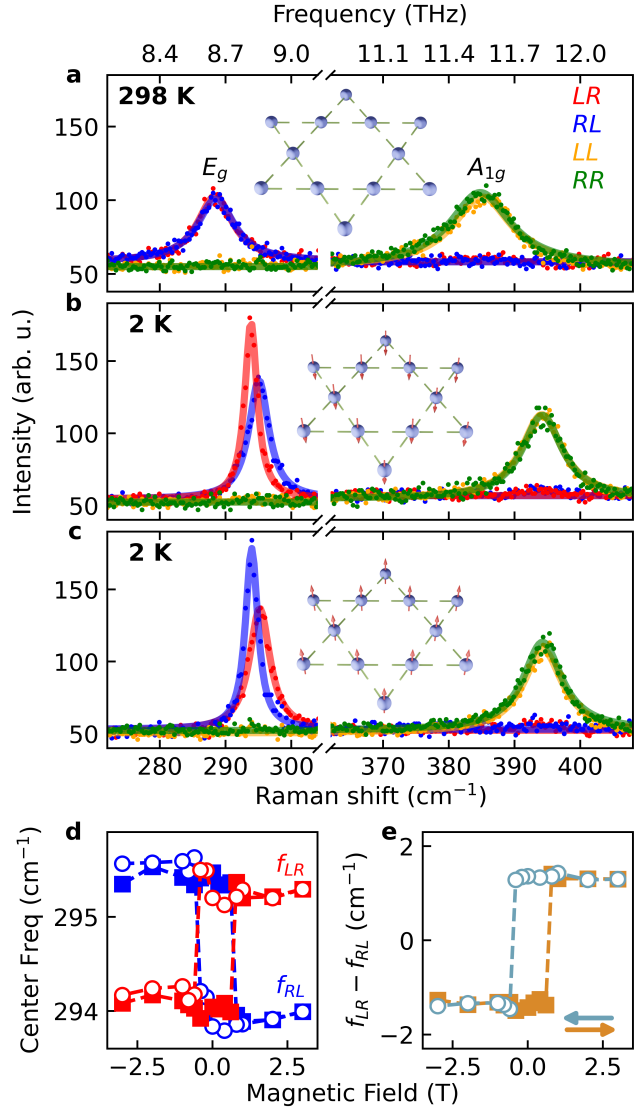

**Fig. 2 Spontaneous chiral phonon splitting induced by magnetization.** (a) At 298 K, Raman spectra, measured with co-circular (*LL* or *RR*) or cross-circular (*LR* or *RL*) polarization, reveal the $A_{1g}$ and doubly-degenerate $E_g$ modes, respectively. (b) When zero-field cooling the sample to 2 K, the two originally degenerate $E_g$ modes spontaneously split in energy with different peak intensities and linewidths. (c) When the magnetization is switched by the external out-of-plane magnetic field, the sign of the splitting switches. (d) Center frequencies of the *LR* and *RL* configurations ($f_{LR}, f_{RL}$) as a function of the applied field show well-defined hysteresis loops. The measurements were performed at 2 K. (e) The frequency splitting, $f_{LR} - f_{RL}$, versus the applied field. The arrows indicate the magnetic field sweep directions. arb. u. stands for arbitrary units.

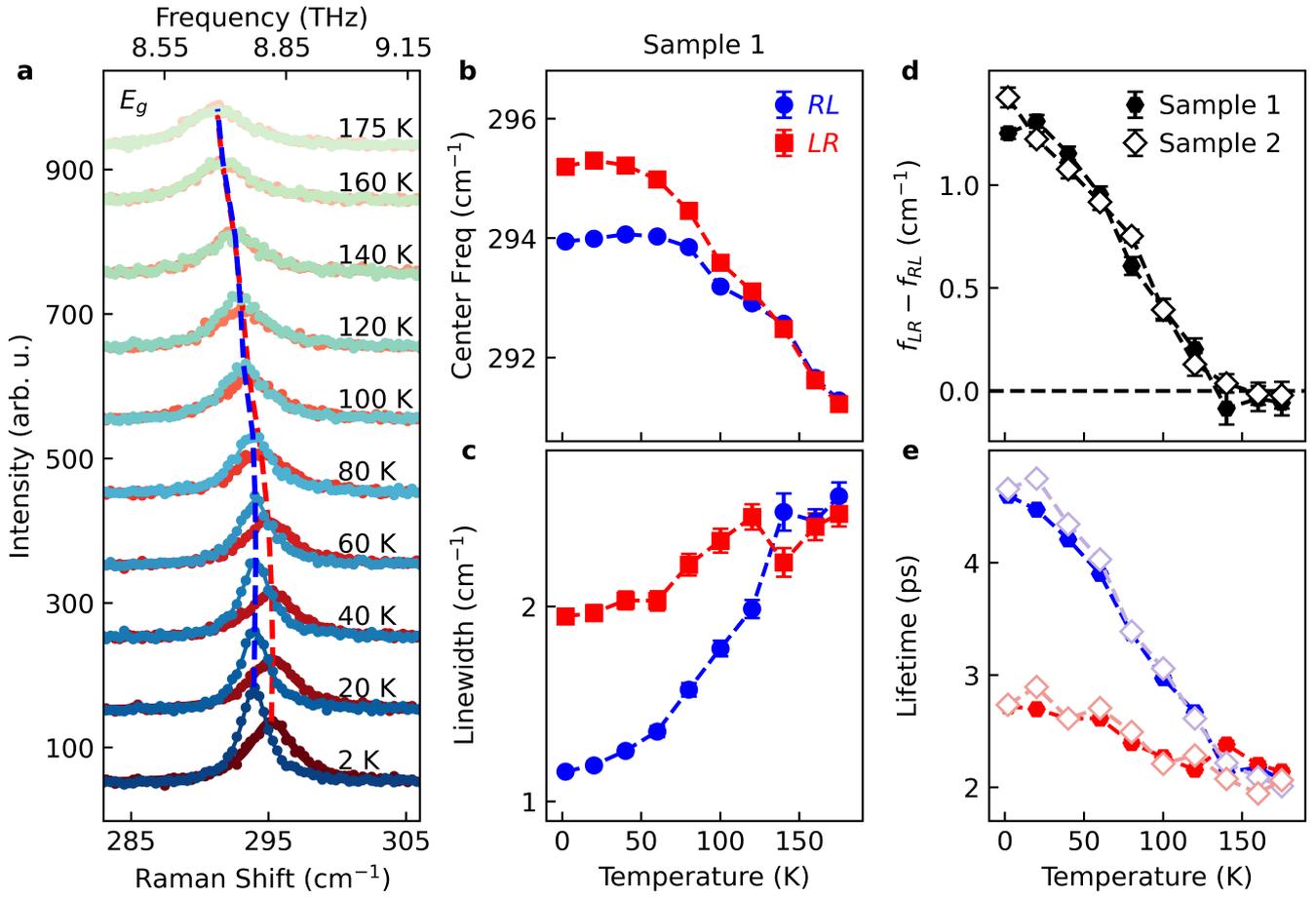

**Fig. 3 Temperature dependence of the chiral phonon modes.** (a) Raman spectrum of the $E_g$ modes for a variety of temperatures at zero applied field. The splitting diminishes with increasing temperature. (b,c) The center frequencies and linewidths obtained by fitting the raw data to the Lorentzian function in Sample 1. From the extracted parameters, the energy splitting and lifetimes are calculated and plotted in (d,e) for both samples. arb. u. stands for arbitrary units.

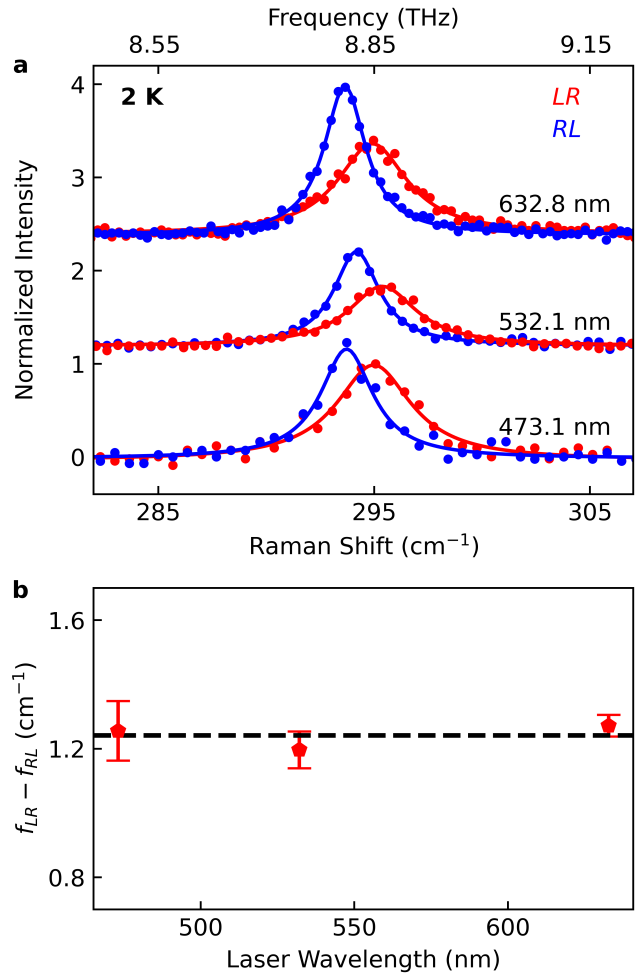

**Fig. 4 Wavelength dependence of the chiral phonon modes.** (a) Raman spectra of the $E_g$ modes at various probe laser wavelengths at 2 K at zero applied field. (b) Frequency splitting versus probe laser wavelength. The black dashed line is the average splitting value for the three probe wavelengths. The lack of wavelength dependence indicates that the chiral phonons split spontaneously at the Γ point.